\documentclass[twocolumn]{aastex6}

\usepackage{natbib}
\usepackage{nicefrac}
\usepackage{amsmath}

\hypersetup{citecolor=blue,linkcolor=blue,urlcolor=blue}
\usepackage[hang]{footmisc}

\renewcommand{\footnotesize}{\scriptsize}
\renewcommand{\textsc}{}

\def\Msun{{\rm M}_\odot}

\shorttitle{{\sc Early Emission From the GW170817 Kilonova}}
\shortauthors{{\sc Arcavi}}

\begin{document}

\title{The First Hours of the GW170817 Kilonova and the Importance of Early Optical and Ultraviolet Observations for Constraining Emission Models}

\author{
Iair~Arcavi\altaffilmark{1,2,3,4}
}

\affil{
\altaffilmark{1}{Department of Physics, University of California, Santa Barbara, CA 93106-9530, USA; \href{mailto:arcavi@ucsb.edu}{arcavi@ucsb.edu}}\\
\altaffilmark{2}{Las Cumbres Observatory, 6740 Cortona Drive, Suite 102, Goleta, CA 93117-5575, USA}\\
\altaffilmark{3}{Kavli Institute for Theoretical Physics, University of California, Santa Barbara, CA 93106, USA}\\
\altaffilmark{4}{Einstein Fellow}\\
}

\begin{abstract}
The kilonova associated with GW170817 displayed early blue emission, which has been interpreted as a signature of either radioactive decay in low-opacity ejecta, relativistic boosting of radioactive decay in high-velocity ejecta, the cooling of material heated by a wind or by a ``cocoon'' surrounding a jet, or a combination thereof. Distinguishing between these mechanisms is important for constraining the ejecta components and their parameters, which tie directly into the physics we can learn from these events. I compile published ultraviolet, optical, and infrared light curves of the GW170817 kilonova and examine whether the combined data set can be used to distinguish between early-emission models. The combined optical data show an early rise consistent with radioactive decay of low-opacity ejecta as the main emission source, but the subsequent decline is fit well by all models. A lack of constraints on the ultraviolet flux during the first few hours after discovery allows for both radioactive decay and other cooling mechanisms to explain the early bolometric light curve. This analysis demonstrates that early (few hours after merger) high-cadence optical and ultraviolet observations will be critical for determining the source of blue emission in future kilonovae.

\end{abstract}

\section{Introduction}

The first detection of gravitational waves from the merger of two neutron stars \citep[GW170817;][]{LIGO170817} was followed by the detection of electromagnetic emission from the same source \citep[see][and references therein]{LIGOmma}. This emission, which spanned the $\gamma$-ray, X-ray, ultraviolet, optical, infrared, and radio wavelengths, can be used to constrain models in a variety of physical regimes. 

The optical-infrared flare following a neutron star merger, known as a kilonova, was theoretically predicted to be powered mainly from the radioactive decay of heavy elements formed in the merger \citep[e.g.][]{Li1998, Rosswog2005, Metzger2010, Roberts2011}. 

During the final coalescence, approximately $10^{-4}$--$10^{-2}\Msun$ of neutron-rich material is predicted to be released at high velocities ($0.1$--$0.3c$) from tidal tails in the equatorial plane and possibly also ejected from the contact region between the two neutron stars in the polar direction \citep[e.g.][]{Hotokezaka2013}. Additional mass could also be ejected in neutrino-driven winds and/or in outflows from a newly formed accretion disk \citep[e.g.][]{Metzger2008, Metzger2018, Grossman2014}. 

The conditions in some or all of these ejecta components could accommodate $r$-process nucleosynthesis of heavy elements, which would radiate as they decay. The detailed emission properties depend on the ejecta mass, velocity, and composition. Heavier elements known as lanthanides (formed in low electron-fraction material) can increase the ejecta opacity by several orders of magnitude \citep{Kasen2013, Tanaka2013}, making the light curve fainter, redder, and longer-lived \citep{Barnes2013, Grossman2014}. 

Each of the ejecta components mentioned above could have a different mass, velocity, and composition, leading to several emission components. The tidal ejecta are expected to be lanthanide-rich and thus produce an approximately week-long infrared transient, while the polar ejecta may have a higher electron fraction, resulting in lanthanide-poor ejecta and a faster and bluer transient. The emission from wind or disk outflows may be either red or blue, depending on the nature and lifetime of the merger product. Even a short-lived ($\sim100$\,ms) massive neutron star could induce enough neutrino irradiation to increase the electron fraction of the wind ejecta and shift its associated emission to the blue (see \citealt{Metzger2017summary} for a review).

Broadly speaking, the properties of the tidal ejecta are most sensitive to the mass ratio of the neutron stars, the properties of the polar ejecta are sensitive to the neutron star radii, and the properties of the wind ejecta are sensitive to the nature of the merger product \citep[][and references therein]{Metzger2017summary}. It is therefore desirable to identify the different emission components in observed kilonovae in order to constrain the properties of their associated ejecta components. 

For example, determining the neutron star masses and radii, as well as the nature of the merger product, can provide novel constraints on the neutron star equation of state. In addition, identifying the existence of a polar component can constrain the viewing angle, which in turn constrains the distance of the merger through the gravitational-wave signal, improving the use of such mergers as cosmological distance probes \citep{LIGOhubble}.

Specifically, it is useful to determine if and how much of the early blue emission seen in the GW170817 kilonova indeed originates from radioactive decay in dynamical polar low-opacity ejecta, in highly relativistic ejecta, in ejecta heated by winds, and/or by the cooling of shock-heated ejecta interacting with a ``cocoon'' produced by a jet created in the merger. This determination could potentially also constrain the uncertainties in the formation and ejection of such a jet \citep[e.g.][]{Murguia-Berthier2014,Murguia-Berthier2017}.

One of the main differences in predictions between radioactive decay vs. cooling from the various other scenarios is the rise time of the emission. Radioactive emission could have an approximately day-long rise due to the requirement that it thermalize and make its way through the expanding ejecta \citep[e.g.][]{Metzger2017}:
\begin{equation}\label{eq:t_rise}
\small
t_{\rm rise}\approx1.6\,{\rm day}\left(\frac{M}{10^{-2}\Msun}\right)^{1/2}\left(\frac{v}{0.1c}\right)^{-1/2}\left(\frac{\kappa}{1\,{\rm cm}^2\,{\rm g}^{-1}}\right)^{1/2}.
\end{equation}
Shock cooling emission, on the other hand, will be essentially instantaneously declining \citep[e.g.][]{Piro2017}. 

An important caveat to Equation \ref{eq:t_rise} is that it assumes a single central heating source diffusing through a uniformly expanding ejecta. The rising behavior of a light curve produced from multiple such components \citep[e.g.][]{Villar2017} or from models with different assumptions \citep[e.g.][]{Kasen2017, Kasliwal2017,Waxman2017,Metzger2018} may differ. Still, given the early and dense coverage of the GW170817 kilonova, examining its rise time may indicate a preference for certain models over others.

The GW170817 kilonova was discovered by telescopes in Chile $\sim11$ hours after the merger, and was only visible for approximately 1--2 hours per night. Therefore, discerning a light curve rise on a $\lesssim24$ hour timescale requires combining observations from multiple sites. This was initially done by \cite{ArcaviGCN} using the Las Cumbres Observatory global network of telescopes, finding an approximately $1$ day rise in single-band observations taken between Chile and Australia. 

Here, I compile the published ultraviolet, optical, and infrared data of the GW170817 kilonova\footnote{named AT 2017gfo \citep{Coulter2017TNS}, SSS17a \citep{Coulter2017}, and DLT17ck \citep{Valenti2017}} obtained from various sources at various observing sites to better constrain the light curve rise and to test whether it can be used to distinguish between the different models of the early blue emission.

\section{Observations}

\begin{figure*}
\includegraphics[width=\textwidth]{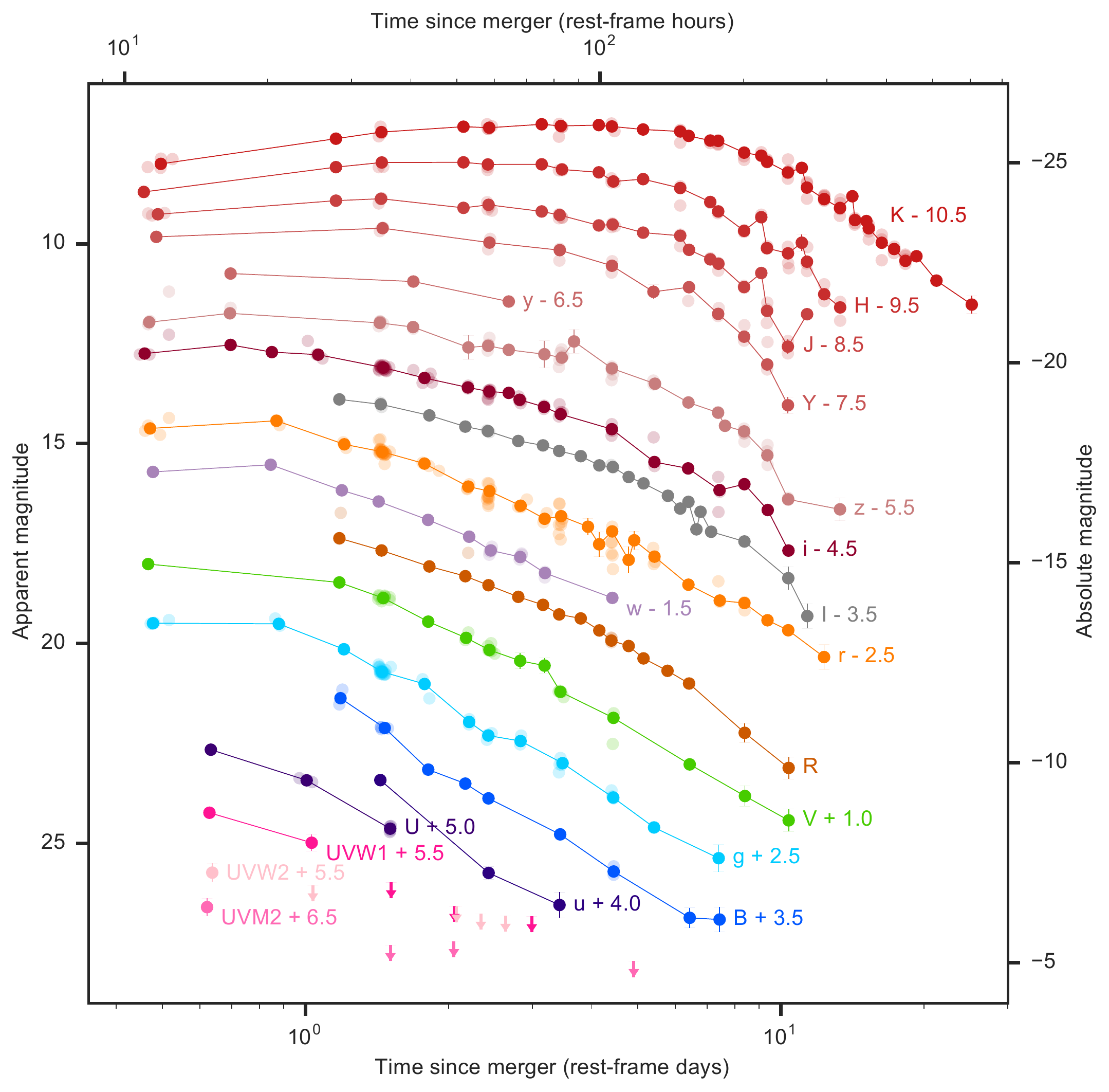}
\caption{\label{fig:lc}Combined ultraviolet-optical-infrared light curve of the GW170817 kilonova with the original data in semi-transparent circles and the binned data in opaque circles (see the text for references). A rise in the optical bands is apparent on a $\sim1$ day timescale. $1\sigma$ error bars are plotted for the binned data and are sometimes smaller than the markers used.}
\end{figure*}

\begin{figure*}[t!]
\includegraphics[width=\textwidth]{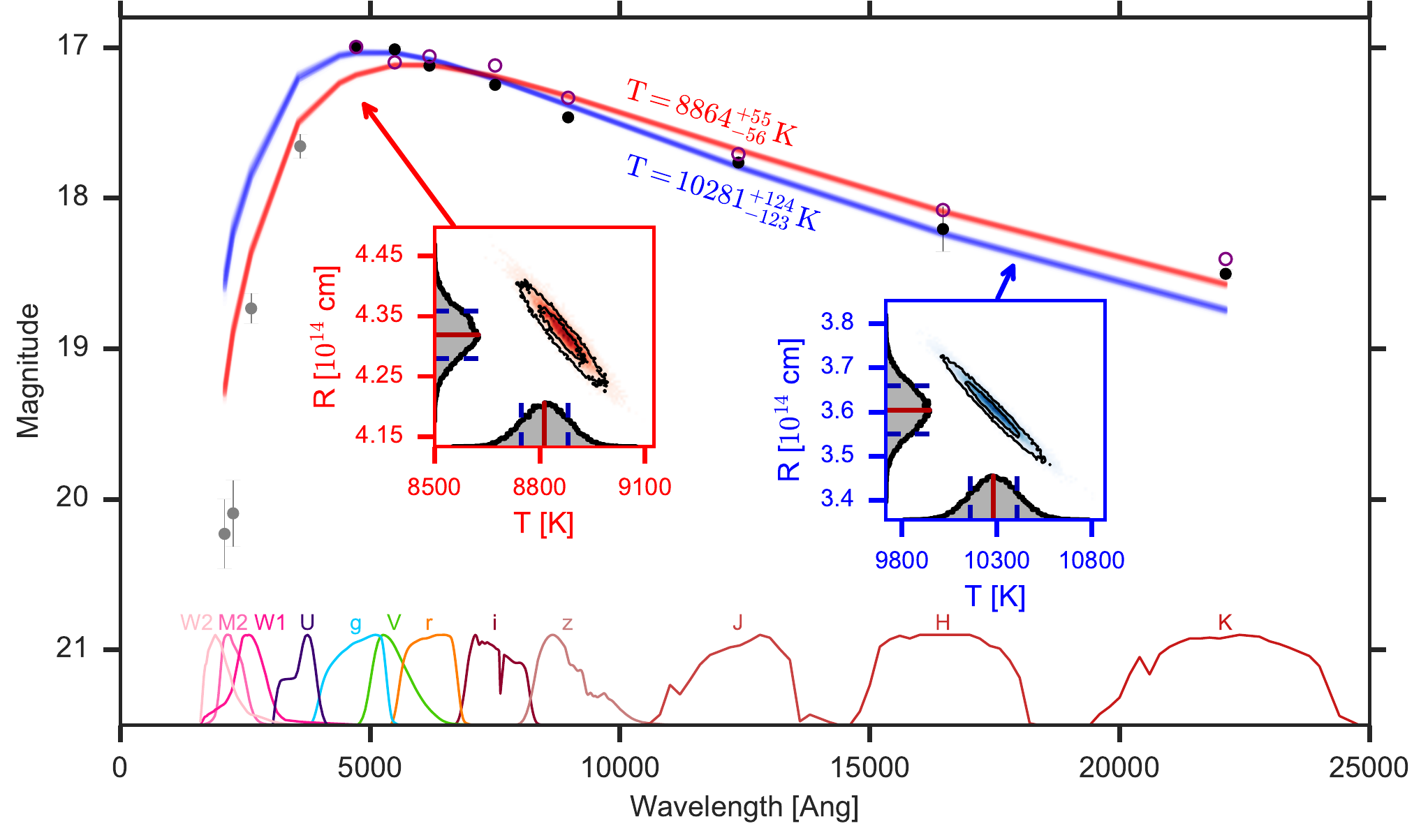}
\caption{\label{fig:sed}Blackbody fits to the optical and infrared SED at discovery (black points; blue fit) and to the ultraviolet, optical, and infrared (the ultraviolet data are from four hours after discovery; gray points; red fit). Including the ultraviolet data reduces the inferred blackbody temperature by $\sim1500$\,K, resulting in a lower bolometric luminosity at discovery (Fig. \ref{fig:ltr}). Optical and infrared magnitudes interpolated to the ultraviolet epoch are shown in purple empty circles for comparison and demonstrate the rapid evolution of the SED at these times. The insets show the corner plots for each MCMC fit, with contour lines denoting 50\% and 90\% bounds, red and blue solid lines (overplotted on each histogram) denoting the mean and median of each parameter distribution (on top of each other in most cases), and dashed lines denoting 68\% confidence bounds. The bandpasses of the different filters used are shown at the bottom of the main plot.}
\end{figure*}

I compile the data from \cite{Andreoni2017}, \cite{Arcavi2017nature}, \cite{Cowperthwaite2017}, \cite{Coulter2017}, \cite{Diaz2017}, \cite{Drout2017}, \cite{Evans2017}, \cite{Hu2017}, \cite{Kasliwal2017}, \cite{Lipunov2017}, \cite{Pian2017}, \cite{Shappee2017}, \cite{Smartt2017}, \cite{Tanvir2017}, \cite{Troja2017}, \cite{Utsumi2017}, \cite{Valenti2017}, and \cite{Pozanenko2018} through the Open Kilonova Catalog\footnote{\url{http://www.kilonova.space}} \citep{Guillochon2017}. Data points with reported errors larger than 0.35 mag are omitted. I correct the data for Milky Way extinction using the \cite{Schlafly2011} extinction maps retrieved via the NASA Extragalactic Database (NED)\footnote{\url{http://ned.ipac.caltech.edu/}}. In the following I assume a distance modulus of $32.98$ and a distance of $39.5$\,Mpc to the host galaxy of the GW170817 kilonova \citep[][retrieved via NED]{Freedman2001}. 

Data from the same filter taken within $0.1$ days of each other are binned together (given the limited observability of the kilonova from ground-based sites, data in most bins were taken within the same half hour). The binning was performed in flux space by weight-averaging the data points according to their reported uncertainties. The original and binned data are presented in Figure \ref{fig:lc}. Though some data points are outliers to their bin, they do not skew the bin average and therefore are not omitted. 

All data points in the Open Kilonova Catalog that include magnitude system information were reported in the AB system, except for a few $V$-band data points that were reported in the Vega system. Since the $V$-band differences between Vega and AB are negligible ($\sim0.02$ mag), I consider all data points to be in the AB system. A definitive compilation of the data will require re-processing all images using homogeneous procedures for host-galaxy light removal, photometry extraction, and magnitude calibration, and is beyond the scope of this Letter.

The combined photometry supports a $\sim1$ day rise in the optical light curves. \cite{ArcaviGCN} reported a rise in the $w$-band - with the combined data, the rise can be seen also in the $r$ and $i$-band data, and at least a flattening can be inferred in the $g$ and $V$ bands on the same time scale. The first ultraviolet observations were obtained four hours after the first optical ones and seem to be already declining at that time.

I construct a bolometric light curve by dividing the combined multi-band data set (excluding the $w$, $y$ and $Y$ bands)\footnote{$w$ band is excluded because it is a broad filter that encompasses the $g$, $r$ and $i$ bands that were observed separately; the $y$ and $Y$ bands are excluded due to differences in the filter characteristics used by the different sources - the flux in these infrared bands is of little consequence when constraining the blue emission mechanism.} into $0.2$ day long epochs, starting from the third epoch. The first epoch is analyzed twice, due to the uncertainty in the ultraviolet flux at discovery (see below). For the second epoch, the $U$ and $UVW1$-band data are interpolated to match the time of optical and infrared observations. 

All epochs with data in at least four different bands are fit to a blackbody using Markov Chain Monte Carlo (MCMC) simulations implemented through the \texttt{emcee} package \citep{Foreman-Mackey2013} to determine the blackbody temperature and radius. The model blackbody SEDs are convolved with the observed filters for each epoch and compared to the observed magnitudes. This method implicitly takes the characteristics of the filter transmissions into account, including the optical tails known as the ``red leaks'' in the $UVW1$ and $UVW2$ filters on board the {\it Neil Gehrels Swift Observatory} Ultraviolet Optical Telescope.

A large uncertainty in the bolometric luminosity in the first epoch is due to the fact that ultraviolet images were obtained four hours after the first optical and infrared ones - a delay comparable to the predicted evolution time scale of the emission at that time. Therefore the bolometric luminosity at the epoch is calculated in two different ways:
\begin{enumerate}
\item By fitting a blackbody only to the simultaneous optical and infrared data (blue fit in Figure \ref{fig:sed}).
\item By fitting a blackbody to the optical, infrared, and (four-hour later) ultraviolet data (red fit in Figure \ref{fig:sed}).
\end{enumerate}

Both fits for the discovery epoch are presented in Figure \ref{fig:sed}. They each match the data well and converge cleanly on a set of blackbody parameters. Including the ultraviolet data from four hours after discovery in the discovery epoch lowers the best-fit temperature by about $1500$\,K, compared to not having any ultraviolet constraints. 

\begin{figure}
\includegraphics[width=\columnwidth]{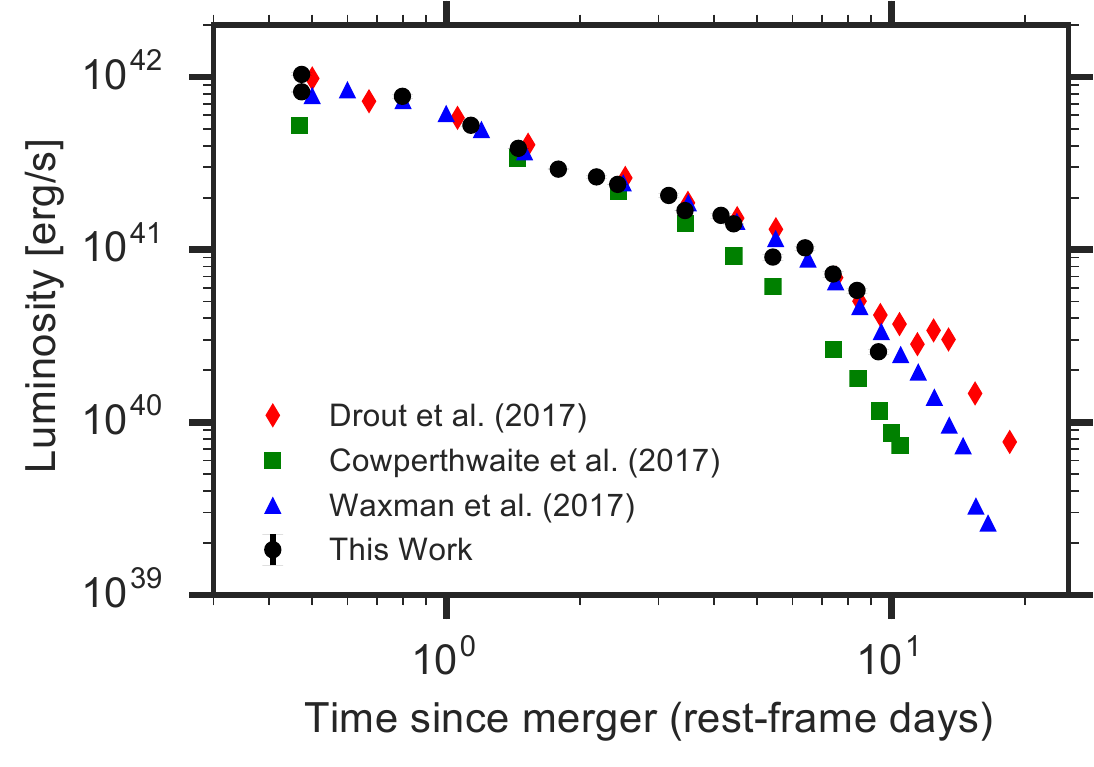}
\caption{\label{fig:lbol}Bolometric light curve of the GW170817 kilonova derived here compared to those from \cite{Cowperthwaite2017}, \cite{Drout2017}, and \citet[][SED integration method]{Waxman2017}.}
\end{figure}

A total of 21 additional epochs are fit. All fits converge onto single-peaked radius and temperature distributions except in five epochs where a small number of walkers converged on disjointed local likelihood maxima, leading to a $>15\%$ relative error in the radius, temperature, or deduced bolometric luminosity. These epochs were removed. The resulting bolometric light curve is presented and compared to other published light curves in Figure \ref{fig:lbol}.

After the first day, the bolometric light curve declines at an initial rate of ${\sim}t^{-1}$ followed by a steeper ${\sim}t^{-1.3}$ decline at later times, consistent with the expected radioactive heating rate.

It is possible that the true SED of the kilonova cannot be described by a single blackbody in all epochs. However, the good fits in most epochs suggest that the blackbody approximation is reasonable. In addition, the bolometric values derived here are almost identical to those derived by performing trapezoidal integration of the multi-band photometric data (without a blackbody assumption) in \cite{Waxman2017}. 

\section{Models}

\begin{figure*}
\centering
\includegraphics[width=\columnwidth]{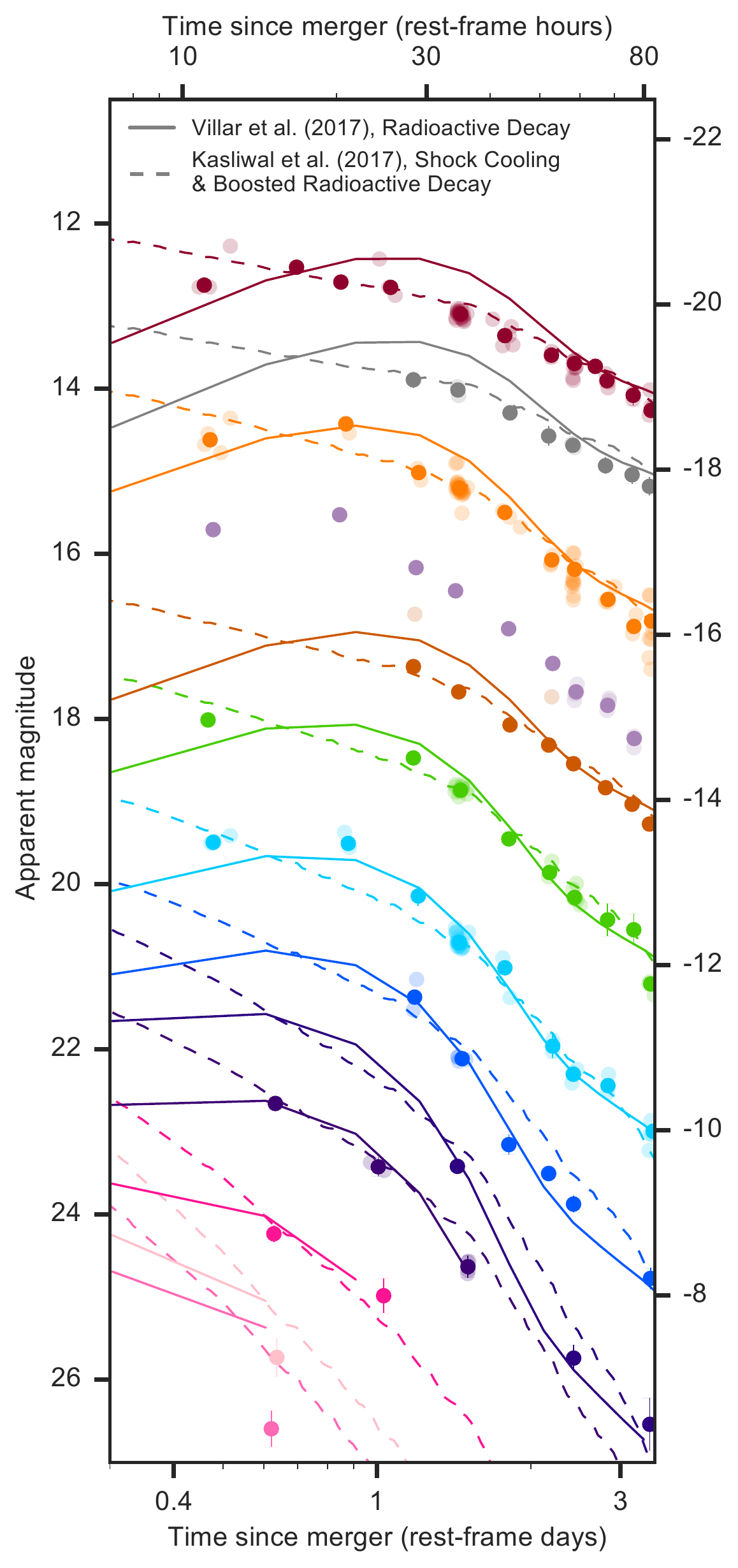}\,\,\,\,\,\,\,\includegraphics[width=\columnwidth]{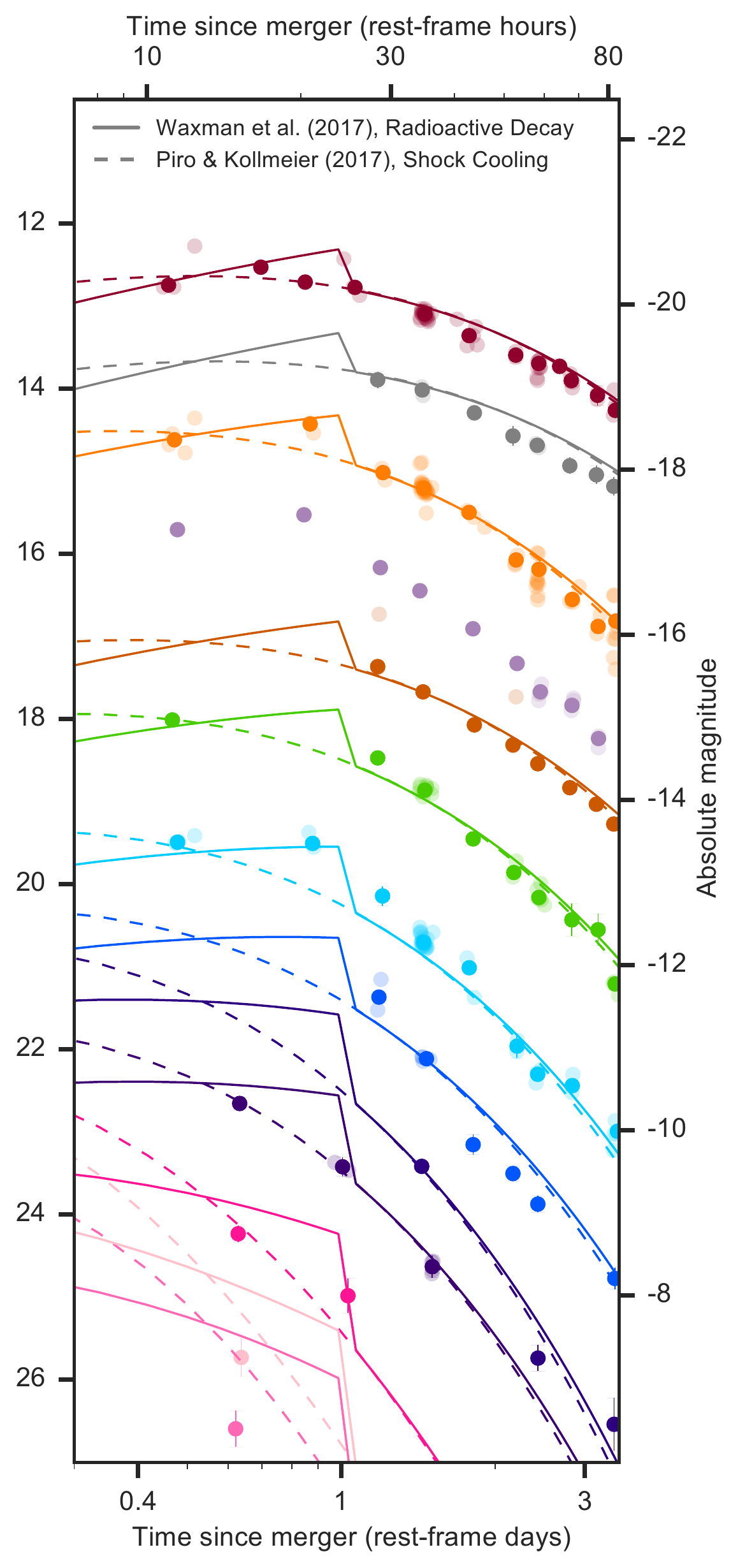}
\caption{\label{fig:lc_models}Optical and ultraviolet light curves for the first 3 days after merger. Left: compared to the radioactive decay luminosity model from \cite{Villar2017} and the combined shock cooling - boosted radioactive decay model from \cite{Kasliwal2017}. Right: compared to the single-component radioactive decay luminosity model from \cite{Waxman2017} and the shock cooling model from \cite{Piro2017}. The \cite{Villar2017} pure radioactive decay model is more consistent with the $\sim1$ day rise seen in the optical bands, but the other models are more consistent with the subsequent decline. The \cite{Waxman2017} model is consistent with both the rise and decline but introduces a sharp jump in the light curves, which in reality may be smoother. Observations at earlier times, where the models differ more substantially, could provide stronger constraints for future events. Colors and filter shifts are the same as in Figure \ref{fig:lc}.}
\end{figure*}

Five models are considered: the analytical multi-component purely radioactive emission model used in \cite{Villar2017}, the analytical single-component purely radioactive emission model of \cite{Waxman2017}, the numerical shock cooling plus boosted radioactive emission model used in \cite{Kasliwal2017}, the analytical purely shock cooling emission model of \cite{Piro2017}, and the \cite{Metzger2018} model considering the cooling of wind-heated ejecta and radioactive emission.

\cite{Villar2017} employ a model that combines a parameterization of the radioactive heating rate from \cite{Korobkin2012}, an analytical approximation to the thermalization fraction from \cite{Barnes2016}, and the diffusion formalism outlined in \cite{Arnett1982} for central energy deposition in a sphere undergoing homologous expansion. A blackbody is assumed with a photospheric radius expanding at a constant velocity until a temperature floor (which is left as a free parameter) is reached. \cite{Villar2017} find a good match to the multi-band light curve combined from various sources (similar to the one produced here) by using a spherical model with three ejecta components, which they denote as blue, purple and red. Each component has a fixed opacity of $\kappa_{\rm blue}=0.5$\,cm$^2$\,g$^{-1}$, $\kappa_{\rm purple}=3$\,cm$^2$\,g$^{-1}$ and $\kappa_{\rm red}=10$\,cm$^2$\,g$^{-1}$. The model has a total of 10 free parameters: 3 ejecta masses, 3 velocities, and 3 temperatures, and a scatter term. Their fitted ejecta masses and ejecta velocities are $M_{\rm blue}=0.02\Msun$, $M_{\rm purple}=0.05\Msun$, $M_{\rm red}=0.01\Msun$, and $v_{\rm blue}=0.27c$, $v_{\rm purple}=0.15c$, $v_{\rm red}=0.14c$ respectively.

\cite{Waxman2017} use a similar literature-combined light curve, but explain it with a single ejecta mass component, parameterized with a power-law velocity distribution, a uniform radioactive energy release rate, and a time-dependent opacity, with the energy deposition governed by electrons and positrons rather than neutrinos and $\gamma$-rays. The model has eight main parameters. Though some of them might be constrained from a priori theoretical considerations, here I consider them free parameters: the normalizations and power-law indices for the velocity, heating rate, and opacity ($v_M$,$\alpha$,$\dot{\varepsilon}_M$,$\beta$, and $\kappa_M$,$\gamma$ respectively), the total ejecta mass $M$, and the effective electron opacity $\kappa_e$. \cite{Waxman2017} find good matches to the bolometric and multi-band light curves with an ejecta velocity gradient of $v\approx0.1c$--$0.3c$, a relatively high ejecta mass of $M\approx0.05\Msun$ and an opacity of $\kappa\approx0.3$\,cm$^2$\,g$^{-1}$ at early times and $\kappa\approx1$\,cm$^2$\,g$^{-1}$ by day 6.

\cite{Kasliwal2017} consider the formation of a cocoon as a newly launched jet interacts with the material ejected by the merger. They use numerical simulations to show that regardless of the fate of the jet, the cocoon can shock and heat the ejecta at relatively large radii compared to the size of the neutron stars. As the ejecta cool they produce blue emission on a time-scale of a few hours \citep{Gottlieb2018}, after which the emission source transitions from shock cooling dominated to radioactive decay dominated, and the main contribution to the blue emission comes from time-dilated and doppler-boosted radioactive decay in high-velocity ($>0.4c$) ejecta.

\cite{Piro2017} are able to match the early light curves of \cite{Drout2017} with shock cooling emission alone. They adapt the analytical shock cooling models from \cite{Nakar2010}, \cite{Piro2013}, and \cite{Nakar2014} to the case of ejecta shocked through energy deposited by a jet \citep{Nakar2017}. The energy deposition is taken to be a power law in velocity, and an expanding blackbody photosphere is assumed. The model has five free parameters: the power-law index $s$, the initial radius where the shock heating was deposited $R$,  the ejecta mass $M_{\rm ej}$, the minimum expansion velocity $v$, and the opacity $\kappa$.

\cite{Metzger2018} consider both cooling and radioactive decay in a neutrino-heated, magnetically accelerated wind created from a hypermassive neutron star that survives $\sim0.1$--$1$\,s after the merger. As with the cocoon, this mechanism can heat ejecta at relatively large radii. A similar power-law energy distribution as in \cite{Piro2017} is assumed, though with a larger ejecta mass moving at higher velocities.

Some of the debate around the different models focuses on whether their inferred parameters for the ejecta mass, velocity, and opacity, as well as their energetics, are compatible with the gravitational-wave constraints for this event and with our assumptions on neutron star and shock cooling physics. Here, I accept the models as they were presented in their respective papers and test how consistent they are with the combined published light curves of the GW170817 kilonova.

\begin{figure*}[t!]
\includegraphics[width=\textwidth]{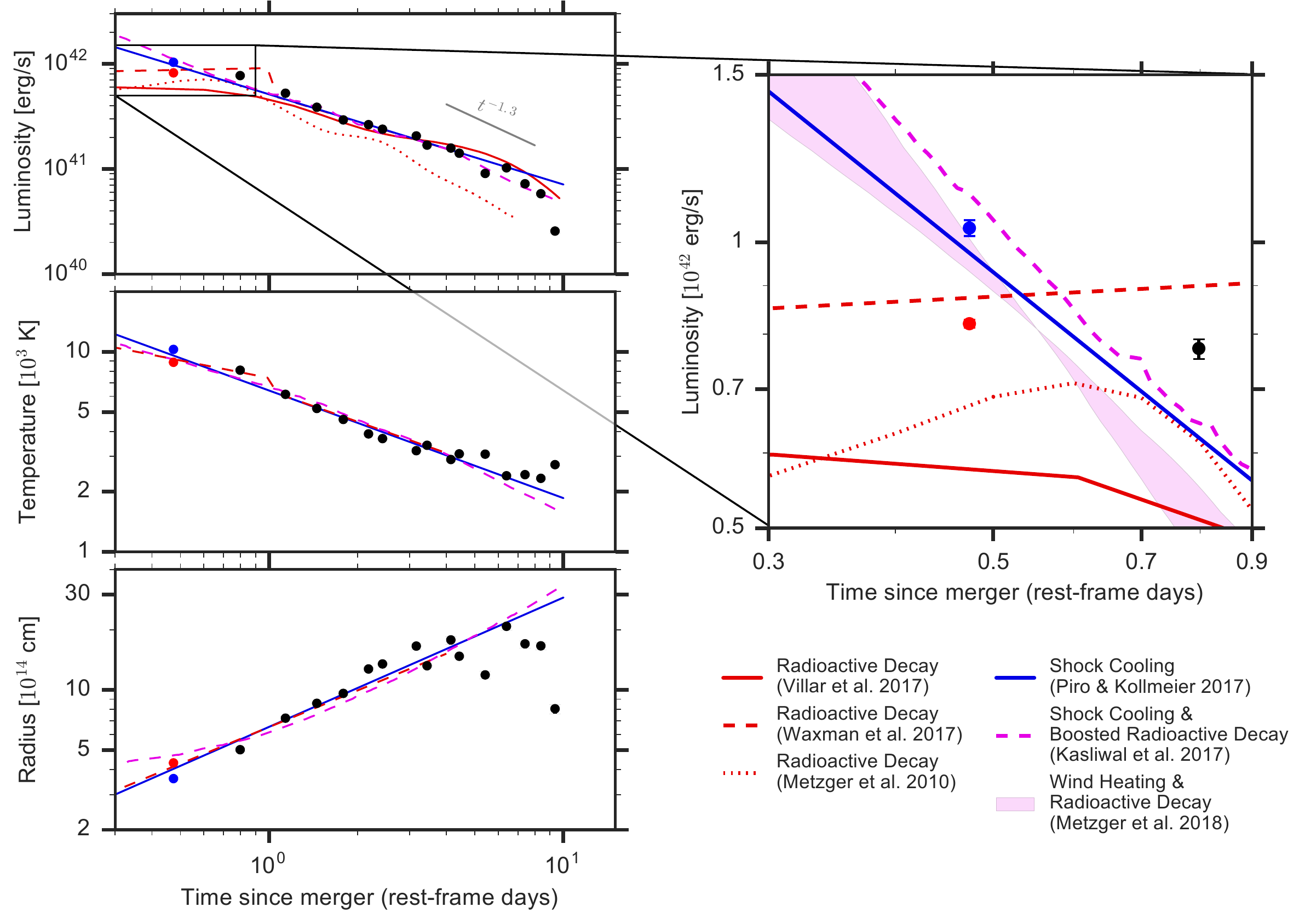}
\caption{\label{fig:ltr}Bolometric luminosity, photospheric temperature, and photospheric radius from the blackbody fits (circles). The results from the two blackbody fits from Figure \ref{fig:sed} are shown for the first epoch: without any assumptions on the ultraviolet emission at that epoch (blue circle), and including the ultraviolet data from four hours later (red circle). That difference is enough to change the early behavior of the bolometric light curve from a shallow slope, resembling that of the pure radioactive emission models (red lines) to a steep slope similar to that of the shock heating, wind heating, and boosted radioactive decay models (blue lines and shaded region). A radioactive heating rate of $t^{-1.3}$ is also shown for comparison to the late bolometric evolution (gray line). (The data used to create this figure are available.)}
\end{figure*}

\section{Analysis}

\subsection{Comparing Models to the Multi-band Light Curves} 

The \cite{Villar2017} model was fit by them to a combined light curve similar to the one presented here, so I take their best fit as-is for comparison. The \cite{Kasliwal2017} model is numerical so I also take their fit as is. The \cite{Piro2017} and \cite{Waxman2017} models, on the other hand, were not formally fit to the data, so I fit them here to the combined light curve using MCMC simulations in a similar way as described for the blackbody fits above (also excluding the $w$-, $y$-, and $Y$-band data).  

The \cite{Piro2017} shock cooling model is fit to the first four days of the combined multi-band light curve, producing a best fit with $s=2.7$, $R=7.1\times10^{10}$\,cm, $M_{\rm ej}=4\times10^{-3}\Msun$, $v=0.2c$, and $\kappa=0.8$\,cm$^2$\,g$^{-1}$. 

The \cite{Waxman2017} model is fit to the full length of the combined multi-band light curve. I take their $f_{\nu\gamma}$ (the energy fraction carried by neutrinos and $\gamma$-rays) to be $0$, which is consistent with their assumption that the dominant energy deposition is governed by electrons and positrons (in any case, this parameter is fully degenerate with $\dot{\varepsilon}_M$). The best fit is obtained with $v_M = 0.15c$, $\alpha = 0.6$, $\dot{\varepsilon}_M = 7.4\times10^9$\,erg\,g$^{-1}$\,s$^{-1}$, $\beta = 0.9$, $\kappa_M = 0.4$\,cm$^2$\,g$^{-1}$, $\gamma = 0.6$, $M = 3.5\times10^{-2}\,\Msun$, and $\kappa_e =0.6$\,cm$^2$\,g$^{-1}$. These values are in rough agreement with the values predicted from theoretical considerations in \cite{Waxman2017}. However, since some of these parameters are likely to be physically correlated \citep[e.g.][]{Barnes2016}, allowing them all to vary freely may be somewhat unphysical.

Figure \ref{fig:lc_models} shows my best-fit \cite{Piro2017} and \cite{Waxman2017} models, as well as the best-fit model from \cite{Villar2017}, and the model of \citet[][I assume blackbody emission to deduce the multi-band magnitudes]{Kasliwal2017}, compared to the ultraviolet and optical data from the first three days. Since the goal here is to perform a qualitative comparison between the model rise and decline behaviors and the combined data, I do not present goodness of fit comparisons.

Instead, it can be seen that all models qualitatively fit the data well. The longer rise seen in the \cite{Villar2017} radioactive model is more consistent with the rise seen in the data, but the shock cooling and boosted emission models match the subsequent declining phase better. The \cite{Waxman2017} model reproduces both the rise and decline phases well, though the jump in their light curve model may be smoother in reality. Since the models differ more at earlier times, stronger constraints could be obtained in future cases with observations taken $<10$ hours after the merger.

The multi-band behavior of each model depends on assumptions about the nature and evolution of their spectra. Additional insights might be gained by comparing the model bolometric predictions to the bolometric light curve.

\subsection{Comparing Models to the Bolometric Light Curve} 

The bolometric luminosity of the same models presented in Figure \ref{fig:lc_models} is compared to the measured bolometric light curve in Figure \ref{fig:ltr}. The effective temperature and photospheric radius of models with single-component values for these parameters are also compared to the inferred values from observations. 

In all cases, the difference between the two blackbody fits considered for the first epoch SED is enough to accommodate all models. The uncertainty in the ultraviolet flux at discovery influences the bolometric light curve slope at discovery from being rapidly declining, consistent with the added heating models, to being slowly declining, consistent with purely radioactive emission models. The radioactive decay model of \cite{Metzger2010}, which has a rise in the bolometric light curve at early epochs, is also roughly consistent with the data. 

It is thus not possible to distinguish between the models even with the bolometric light curve, given the uncertainty in the early ultraviolet behavior. 

\section{Conclusions}

By combining the photometry of the GW170817 kilonova from various published datasets, a rise time of $\sim1$ day is confirmed in the optical bands. This rise is better reproduced by radioactive decay in low-opacity ejecta, but models with alternative or additional heating sources and the boosted radioactive decay model all reproduce the subsequent decline as well or better.

Because the first data were obtained just at the cusp of peak optical luminosity, and due to the four-hour lag between the discovery epoch and the first ultraviolet observations, neither of the models tested here can be ruled out. In order to distinguish between such models for the early blue emission in future kilonovae, observations in the optical and ultraviolet bands need to be obtained even earlier (i.e. one to a few hours after the merger). Such early observations could also constrain the contribution of free-neutron decay as an additional source of ($\sim$hour-long) blue emission \citep{Metzger2015}.

Obtaining observations on these time scales relies on the quick availability of the gravitational-wave localization. In the case of GW170817, a five-hour delay in the localization (due to a glitch in one of the detectors) prevented earlier discovery of the kilonova. Had the localization been available even one hour earlier, the Las Cumbres Observatory follow-up program would have detected the kilonova from South Africa, five hours before it was discovered over Chile \citep{Arcavi2017nature}. {\it Swift} might have also detected it earlier. Such observations should therefore be possible for future events. Future events could also benefit from space-based wide-field ultraviolet imagers, such as the proposed {\it ULTRASAT} mission \citep{Ganot2016}.

\acknowledgments
I am grateful to E.~Nakar, T.~Piro, B.~Metzger, E.~O.~Ofek, G. Hosseinzadeh, D.~A.~Howell, E.~Waxman, and to the participants of the Kavli Institute for Theoretical Physics (KITP) rapid response ``Astrophysics from a Neutron Star Merger'' program for helpful discussions. I also thank V.~A.~Villar, E.~O.~Ofek, O.~Gottlieb, E.~Nakar, and B.~Metzger for providing me with their model light curves and for help interpreting them; to J.~Guillochon for assistance in retrieving data from the Open Kilonova Catalog; and to the anonymous referee for helpful comments. Support for IA was provided by NASA through the Einstein Fellowship Program, grant PF6-170148. The KITP is supported in part by the National Science Foundation under grant No. NSF PHY 17-48958.

\software{\texttt{astropy} \citep{Astropy2013}, \texttt{emcee} \citep{Foreman-Mackey2013}, \texttt{matplotlib} \citep{Hunter2007}, \texttt{numpy} \citep{VanDerWalt2011}, \texttt{pysynphot} \citep{Pysynphot2013}.}


\begin{thebibliography}{}
\expandafter\ifx\csname natexlab\endcsname\relax\def\natexlab#1{#1}\fi
\providecommand{\url}[1]{\href{#1}{#1}}

\bibitem[{{Abbott} {et~al.}(2017){Abbott}, {Abbott}, {Abbott}, {Acernese},
  {Ackley}, {Adams}, {Adams}, {Addesso}, {Adhikari}, {Adya}, \&
  et~al.}]{LIGOhubble}
{Abbott}, B.~P., {Abbott}, R., {Abbott}, T.~D., {et~al.} 2017, \nat, 551, 85

\bibitem[{{Andreoni} {et~al.}(2017){Andreoni}, {Ackley}, {Cooke}, {Acharyya},
  {Allison}, {Anderson}, {Ashley}, {Baade}, {Bailes}, {Bannister}, {Beardsley},
  {Bessell}, {Bian}, {Bland}, {Boer}, {Booler}, {Brandeker}, {Brown},
  {Buckley}, {Chang}, {Coward}, {Crawford}, {Crisp}, {Crosse}, {Cucchiara},
  {Cup{\'a}k}, {de Gois}, {Deller}, {Devillepoix}, {Dobie}, {Elmer}, {Emrich},
  {Farah}, {Farrell}, {Franzen}, {Gaensler}, {Galloway}, {Gendre}, {Giblin},
  {Goobar}, {Green}, {Hancock}, {Hartig}, {Howell}, {Horsley}, {Hotan},
  {Howie}, {Hu}, {Hu}, {James}, {Johnston}, {Johnston-Hollitt}, {Kaplan},
  {Kasliwal}, {Keane}, {Kenney}, {Klotz}, {Lau}, {Laugier}, {Lenc}, {Li},
  {Liang}, {Lidman}, {Luvaul}, {Lynch}, {Ma}, {Macpherson}, {Mao},
  {McClelland}, {McCully}, {M{\"o}ller}, {Morales}, {Morris}, {Murphy},
  {Noysena}, {Onken}, {Orange}, {Os{\l}owski}, {Pallot}, {Paxman}, {Potter},
  {Pritchard}, {Raja}, {Ridden-Harper}, {Romero-Colmenero}, {Sadler}, {Sansom},
  {Scalzo}, {Schmidt}, {Scott}, {Seghouani}, {Shang}, {Shannon}, {Shao},
  {Shara}, {Sharp}, {Sokolowski}, {Sollerman}, {Staff}, {Steele}, {Sun},
  {Suntzeff}, {Tao}, {Tingay}, {Towner}, {Thierry}, {Trott}, {Tucker},
  {V{\"a}is{\"a}nen}, {Krishnan}, {Walker}, {Wang}, {Wang}, {Wayth}, {Whiting},
  {Williams}, {Williams}, {Wolf}, {Wu}, {Wu}, {Yang}, {Yuan}, {Zhang}, {Zhou},
  \& {Zovaro}}]{Andreoni2017}
{Andreoni}, I., {Ackley}, K., {Cooke}, J., {et~al.} 2017, \pasa, 34, e069

\bibitem[{{Arcavi} {et~al.}(2017{\natexlab{a}}){Arcavi}, {Howell}, {McCully},
  {Hosseinzadeh}, {Vasylyev}, {Zalzman}, {Poznanski}, {Singer}, {Valenti},
  {Piran}, {Kasen}, {Barnes}, \& {Fong}}]{ArcaviGCN}
{Arcavi}, I., {Howell}, D.~A., {McCully}, C., {et~al.} 2017{\natexlab{a}}, GCN
  Circ., 21538

\bibitem[{{Arcavi} {et~al.}(2017{\natexlab{b}}){Arcavi}, {Hosseinzadeh},
  {Howell}, {McCully}, {Poznanski}, {Kasen}, {Barnes}, {Zaltzman}, {Vasylyev},
  {Maoz}, \& {Valenti}}]{Arcavi2017nature}
{Arcavi}, I., {Hosseinzadeh}, G., {Howell}, D.~A., {et~al.} 2017{\natexlab{b}},
  \nat, 551, 64

\bibitem[{{Arnett}(1982)}]{Arnett1982}
{Arnett}, W.~D. 1982, \apj, 253, 785

\bibitem[{{Astropy Collaboration} {et~al.}(2013){Astropy Collaboration},
  {Robitaille}, {Tollerud}, {Greenfield}, {Droettboom}, {Bray}, {Aldcroft},
  {Davis}, {Ginsburg}, {Price-Whelan}, {Kerzendorf}, {Conley}, {Crighton},
  {Barbary}, {Muna}, {Ferguson}, {Grollier}, {Parikh}, {Nair}, {Unther},
  {Deil}, {Woillez}, {Conseil}, {Kramer}, {Turner}, {Singer}, {Fox}, {Weaver},
  {Zabalza}, {Edwards}, {Azalee Bostroem}, {Burke}, {Casey}, {Crawford},
  {Dencheva}, {Ely}, {Jenness}, {Labrie}, {Lim}, {Pierfederici}, {Pontzen},
  {Ptak}, {Refsdal}, {Servillat}, \& {Streicher}}]{Astropy2013}
{Astropy Collaboration}, {Robitaille}, T.~P., {Tollerud}, E.~J., {et~al.} 2013,
  \aap, 558, A33

\bibitem[{{Barnes} \& {Kasen}(2013)}]{Barnes2013}
{Barnes}, J., \& {Kasen}, D. 2013, \apj, 775, 18

\bibitem[{{Barnes} {et~al.}(2016){Barnes}, {Kasen}, {Wu}, \&
  {Mart{\'{\i}}nez-Pinedo}}]{Barnes2016}
{Barnes}, J., {Kasen}, D., {Wu}, M.-R., \& {Mart{\'{\i}}nez-Pinedo}, G. 2016,
  \apj, 829, 110

\bibitem[{{Coulter} {et~al.}(2017{\natexlab{a}}){Coulter}, {Kilpatrick},
  {Siebert}, {Foley}, {Shappee}, {Drout}, {Simon}, {Piro}, \&
  {Rest}}]{Coulter2017TNS}
{Coulter}, D.~A., {Kilpatrick}, C.~D., {Siebert}, M.~R., {et~al.}
  2017{\natexlab{a}}, Transient Name Server Discovery Report, 1030

\bibitem[{{Coulter} {et~al.}(2017{\natexlab{b}}){Coulter}, {Foley},
  {Kilpatrick}, {Drout}, {Piro}, {Shappee}, {Siebert}, {Simon}, {Ulloa},
  {Kasen}, {Madore}, {Murguia-Berthier}, {Pan}, {Prochaska}, {Ramirez-Ruiz},
  {Rest}, \& {Rojas-Bravo}}]{Coulter2017}
{Coulter}, D.~A., {Foley}, R.~J., {Kilpatrick}, C.~D., {et~al.}
  2017{\natexlab{b}}, Science, 358, 1556

\bibitem[{{Cowperthwaite} {et~al.}(2017){Cowperthwaite}, {Berger}, {Villar},
  {Metzger}, {Nicholl}, {Chornock}, {Blanchard}, {Fong}, {Margutti},
  {Soares-Santos}, {Alexander}, {Allam}, {Annis}, {Brout}, {Brown}, {Butler},
  {Chen}, {Diehl}, {Doctor}, {Drout}, {Eftekhari}, {Farr}, {Finley}, {Foley},
  {Frieman}, {Fryer}, {Garc{\'{\i}}a-Bellido}, {Gill}, {Guillochon}, {Herner},
  {Holz}, {Kasen}, {Kessler}, {Marriner}, {Matheson}, {Neilsen}, {Quataert},
  {Palmese}, {Rest}, {Sako}, {Scolnic}, {Smith}, {Tucker}, {Williams},
  {Balbinot}, {Carlin}, {Cook}, {Durret}, {Li}, {Lopes}, {Louren{\c c}o},
  {Marshall}, {Medina}, {Muir}, {Mu{\~n}oz}, {Sauseda}, {Schlegel}, {Secco},
  {Vivas}, {Wester}, {Zenteno}, {Zhang}, {Abbott}, {Banerji}, {Bechtol},
  {Benoit-L{\'e}vy}, {Bertin}, {Buckley-Geer}, {Burke}, {Capozzi}, {Carnero
  Rosell}, {Carrasco Kind}, {Castander}, {Crocce}, {Cunha}, {D'Andrea}, {da
  Costa}, {Davis}, {DePoy}, {Desai}, {Dietrich}, {Drlica-Wagner}, {Eifler},
  {Evrard}, {Fernandez}, {Flaugher}, {Fosalba}, {Gaztanaga}, {Gerdes},
  {Giannantonio}, {Goldstein}, {Gruen}, {Gruendl}, {Gutierrez}, {Honscheid},
  {Jain}, {James}, {Jeltema}, {Johnson}, {Johnson}, {Kent}, {Krause}, {Kron},
  {Kuehn}, {Nuropatkin}, {Lahav}, {Lima}, {Lin}, {Maia}, {March}, {Martini},
  {McMahon}, {Menanteau}, {Miller}, {Miquel}, {Mohr}, {Neilsen}, {Nichol},
  {Ogando}, {Plazas}, {Roe}, {Romer}, {Roodman}, {Rykoff}, {Sanchez},
  {Scarpine}, {Schindler}, {Schubnell}, {Sevilla-Noarbe}, {Smith}, {Smith},
  {Sobreira}, {Suchyta}, {Swanson}, {Tarle}, {Thomas}, {Thomas}, {Troxel},
  {Vikram}, {Walker}, {Wechsler}, {Weller}, {Yanny}, \&
  {Zuntz}}]{Cowperthwaite2017}
{Cowperthwaite}, P.~S., {Berger}, E., {Villar}, V.~A., {et~al.} 2017, \apjl,
  848, L17

\bibitem[{{D{\'{\i}}az} {et~al.}(2017){D{\'{\i}}az}, {Macri}, {Garcia Lambas},
  {Mendes de Oliveira}, {Nilo Castell{\'o}n}, {Ribeiro}, {S{\'a}nchez},
  {Schoenell}, {Abramo}, {Akras}, {Alcaniz}, {Artola}, {Beroiz}, {Bonoli},
  {Cabral}, {Camuccio}, {Castillo}, {Chavushyan}, {Coelho}, {Colazo},
  {Costa-Duarte}, {Cuevas Larenas}, {DePoy}, {Dom{\'{\i}}nguez Romero},
  {Dultzin}, {Fern{\'a}ndez}, {Garc{\'{\i}}a}, {Girardini}, {Gon{\c c}alves},
  {Gon{\c c}alves}, {Gurovich}, {Jim{\'e}nez-Teja}, {Kanaan}, {Lares}, {Lopes
  de Oliveira}, {L{\'o}pez-Cruz}, {Marshall}, {Melia}, {Molino}, {Padilla},
  {Pe{\~n}uela}, {Placco}, {Qui{\~n}ones}, {Ram{\'{\i}}rez Rivera}, {Renzi},
  {Riguccini}, {R{\'{\i}}os-L{\'o}pez}, {Rodriguez}, {Sampedro}, {Schneiter},
  {Sodr{\'e}}, {Starck}, {Torres-Flores}, {Tornatore}, \&
  {Zadro{\.z}ny}}]{Diaz2017}
{D{\'{\i}}az}, M.~C., {Macri}, L.~M., {Garcia Lambas}, D., {et~al.} 2017,
  \apjl, 848, L29

\bibitem[{{Drout} {et~al.}(2017){Drout}, {Piro}, {Shappee}, {Kilpatrick},
  {Simon}, {Contreras}, {Coulter}, {Foley}, {Siebert}, {Morrell}, {Boutsia},
  {Di Mille}, {Holoien}, {Kasen}, {Kollmeier}, {Madore}, {Monson},
  {Murguia-Berthier}, {Pan}, {Prochaska}, {Ramirez-Ruiz}, {Rest}, {Adams},
  {Alatalo}, {Ba{\~n}ados}, {Baughman}, {Beers}, {Bernstein}, {Bitsakis},
  {Campillay}, {Hansen}, {Higgs}, {Ji}, {Maravelias}, {Marshall}, {Bidin},
  {Prieto}, {Rasmussen}, {Rojas-Bravo}, {Strom}, {Ulloa},
  {Vargas-Gonz{\'a}lez}, {Wan}, \& {Whitten}}]{Drout2017}
{Drout}, M.~R., {Piro}, A.~L., {Shappee}, B.~J., {et~al.} 2017, Science, 358,
  1570

\bibitem[{{Evans} {et~al.}(2017){Evans}, {Cenko}, {Kennea}, {Emery}, {Kuin},
  {Korobkin}, {Wollaeger}, {Fryer}, {Madsen}, {Harrison}, {Xu}, {Nakar},
  {Hotokezaka}, {Lien}, {Campana}, {Oates}, {Troja}, {Breeveld}, {Marshall},
  {Barthelmy}, {Beardmore}, {Burrows}, {Cusumano}, {D'Ai}, {D'Avanzo},
  {D'Elia}, {de Pasquale}, {Even}, {Fontes}, {Forster}, {Garcia}, {Giommi},
  {Grefenstette}, {Gronwall}, {Hartmann}, {Heida}, {Hungerford}, {Kasliwal},
  {Krimm}, {Levan}, {Malesani}, {Melandri}, {Miyasaka}, {Nousek}, {O'Brien},
  {Osborne}, {Pagani}, {Page}, {Palmer}, {Perri}, {Pike}, {Racusin}, {Rosswog},
  {Siegel}, {Sakamoto}, {Sbarufatti}, {Tagliaferri}, {Tanvir}, \&
  {Tohuvavohu}}]{Evans2017}
{Evans}, P.~A., {Cenko}, S.~B., {Kennea}, J.~A., {et~al.} 2017, Science, 358,
  1565

\bibitem[{{Foreman-Mackey} {et~al.}(2013){Foreman-Mackey}, {Hogg}, {Lang}, \&
  {Goodman}}]{Foreman-Mackey2013}
{Foreman-Mackey}, D., {Hogg}, D.~W., {Lang}, D., \& {Goodman}, J. 2013, \pasp,
  125, 306

\bibitem[{{Freedman} {et~al.}(2001){Freedman}, {Madore}, {Gibson}, {Ferrarese},
  {Kelson}, {Sakai}, {Mould}, {Kennicutt}, {Ford}, {Graham}, {Huchra},
  {Hughes}, {Illingworth}, {Macri}, \& {Stetson}}]{Freedman2001}
{Freedman}, W.~L., {Madore}, B.~F., {Gibson}, B.~K., {et~al.} 2001, \apj, 553,
  47

\bibitem[{{Ganot} {et~al.}(2016){Ganot}, {Gal-Yam}, {Ofek}, {Sagiv}, {Waxman},
  {Lapid}, {Kulkarni}, {Ben-Ami}, {Kasliwal}, {The ULTRASAT Science Team},
  {Chelouche}, {Rafter}, {Behar}, {Laor}, {Poznanski}, {Nakar}, {Maoz},
  {Trakhtenbrot}, {WTTH Consortium}, {Neill}, {Barlow}, {Martin}, {Gezari},
  {the GALEX Science Team}, {Arcavi}, {Bloom}, {Nugent}, {Sullivan}, \&
  {Palomar Transient Factory}}]{Ganot2016}
{Ganot}, N., {Gal-Yam}, A., {Ofek}, E.~O., {et~al.} 2016, \apj, 820, 57

\bibitem[{{Gottlieb} {et~al.}(2018){Gottlieb}, {Nakar}, \&
  {Piran}}]{Gottlieb2018}
{Gottlieb}, O., {Nakar}, E., \& {Piran}, T. 2018, \mnras, 473, 576

\bibitem[{{Grossman} {et~al.}(2014){Grossman}, {Korobkin}, {Rosswog}, \&
  {Piran}}]{Grossman2014}
{Grossman}, D., {Korobkin}, O., {Rosswog}, S., \& {Piran}, T. 2014, \mnras,
  439, 757

\bibitem[{{Guillochon} {et~al.}(2017){Guillochon}, {Parrent}, {Kelley}, \&
  {Margutti}}]{Guillochon2017}
{Guillochon}, J., {Parrent}, J., {Kelley}, L.~Z., \& {Margutti}, R. 2017, \apj,
  835, 64

\bibitem[{{Hotokezaka} {et~al.}(2013){Hotokezaka}, {Kiuchi}, {Kyutoku},
  {Okawa}, {Sekiguchi}, {Shibata}, \& {Taniguchi}}]{Hotokezaka2013}
{Hotokezaka}, K., {Kiuchi}, K., {Kyutoku}, K., {et~al.} 2013, \prd, 87, 024001

\bibitem[{{Hu} {et~al.}(2017){Hu}, {Wu}, {Andreoni}, {Ashley}, {Cooke}, {Cui},
  {Du}, {Dai}, {Gu}, {Hu}, {Lu}, {Li}, {Li}, {Liang}, {Liu}, {Ma}, {Shang},
  {Sun}, {Suntzeff}, {Tao}, {Uddin}, {Wang}, {Wang}, {Wen}, {Xiao}, {Xu},
  {Yang}, {Yang}, {Yuan}, {Zhou}, {Zhang}, {Zhou}, \& {Zhu}}]{Hu2017}
{Hu}, L., {Wu}, X., {Andreoni}, I., {et~al.} 2017, ArXiv e-prints,
  arXiv:1710.05462

\bibitem[{Hunter(2007)}]{Hunter2007}
Hunter, J.~D. 2007, Computing in Science Engineering, 9, 90

\bibitem[{{Kasen} {et~al.}(2013){Kasen}, {Badnell}, \& {Barnes}}]{Kasen2013}
{Kasen}, D., {Badnell}, N.~R., \& {Barnes}, J. 2013, \apj, 774, 25

\bibitem[{{Kasen} {et~al.}(2017){Kasen}, {Metzger}, {Barnes}, {Quataert}, \&
  {Ramirez-Ruiz}}]{Kasen2017}
{Kasen}, D., {Metzger}, B., {Barnes}, J., {Quataert}, E., \& {Ramirez-Ruiz}, E.
  2017, \nat, 551, 80

\bibitem[{{Kasliwal} {et~al.}(2017){Kasliwal}, {Nakar}, {Singer}, {Kaplan},
  {Cook}, {Van Sistine}, {Lau}, {Fremling}, {Gottlieb}, {Jencson}, {Adams},
  {Feindt}, {Hotokezaka}, {Ghosh}, {Perley}, {Yu}, {Piran}, {Allison},
  {Anupama}, {Balasubramanian}, {Bannister}, {Bally}, {Barnes}, {Barway},
  {Bellm}, {Bhalerao}, {Bhattacharya}, {Blagorodnova}, {Bloom}, {Brady},
  {Cannella}, {Chatterjee}, {Cenko}, {Cobb}, {Copperwheat}, {Corsi}, {De},
  {Dobie}, {Emery}, {Evans}, {Fox}, {Frail}, {Frohmaier}, {Goobar}, {Hallinan},
  {Harrison}, {Helou}, {Hinderer}, {Ho}, {Horesh}, {Ip}, {Itoh}, {Kasen},
  {Kim}, {Kuin}, {Kupfer}, {Lynch}, {Madsen}, {Mazzali}, {Miller}, {Mooley},
  {Murphy}, {Ngeow}, {Nichols}, {Nissanke}, {Nugent}, {Ofek}, {Qi}, {Quimby},
  {Rosswog}, {Rusu}, {Sadler}, {Schmidt}, {Sollerman}, {Steele}, {Williamson},
  {Xu}, {Yan}, {Yatsu}, {Zhang}, \& {Zhao}}]{Kasliwal2017}
{Kasliwal}, M.~M., {Nakar}, E., {Singer}, L.~P., {et~al.} 2017, Science, 358,
  1559

\bibitem[{{Korobkin} {et~al.}(2012){Korobkin}, {Rosswog}, {Arcones}, \&
  {Winteler}}]{Korobkin2012}
{Korobkin}, O., {Rosswog}, S., {Arcones}, A., \& {Winteler}, C. 2012, \mnras,
  426, 1940

\bibitem[{{Li} \& {Paczy{\'n}ski}(1998)}]{Li1998}
{Li}, L.-X., \& {Paczy{\'n}ski}, B. 1998, \apjl, 507, L59

\bibitem[{{LIGO Scientific Collaboration} \& {Virgo
  Collaboration}(2017{\natexlab{a}})}]{LIGO170817}
{LIGO Scientific Collaboration}, \& {Virgo Collaboration}. 2017{\natexlab{a}},
  Physical Review Letters, 119, 161101

\bibitem[{{LIGO Scientific Collaboration} \& {Virgo
  Collaboration}(2017{\natexlab{b}})}]{LIGOmma}
---. 2017{\natexlab{b}}, \apjl, 848, L12

\bibitem[{{Lipunov} {et~al.}(2017){Lipunov}, {Gorbovskoy}, {Kornilov},
  {.~Tyurina}, {Balanutsa}, {Kuznetsov}, {Vlasenko}, {Kuvshinov}, {Gorbunov},
  {Buckley}, {Krylov}, {Podesta}, {Lopez}, {Podesta}, {Levato}, {Saffe},
  {Mallamachi}, {Potter}, {Budnev}, {Gress}, {Ishmuhametova}, {Vladimirov},
  {Zimnukhov}, {Yurkov}, {Sergienko}, {Gabovich}, {Rebolo}, {Serra-Ricart},
  {Israelyan}, {Chazov}, {Wang}, {Tlatov}, \& {Panchenko}}]{Lipunov2017}
{Lipunov}, V.~M., {Gorbovskoy}, E., {Kornilov}, V.~G., {et~al.} 2017, \apjl,
  850, L1

\bibitem[{{Metzger}(2017{\natexlab{a}})}]{Metzger2017summary}
{Metzger}, B.~D. 2017{\natexlab{a}}, ArXiv e-prints, arXiv:1710.05931

\bibitem[{{Metzger}(2017{\natexlab{b}})}]{Metzger2017}
---. 2017{\natexlab{b}}, Living Reviews in Relativity, 20, 3

\bibitem[{{Metzger} {et~al.}(2015){Metzger}, {Bauswein}, {Goriely}, \&
  {Kasen}}]{Metzger2015}
{Metzger}, B.~D., {Bauswein}, A., {Goriely}, S., \& {Kasen}, D. 2015, \mnras,
  446, 1115

\bibitem[{{Metzger} {et~al.}(2008){Metzger}, {Piro}, \&
  {Quataert}}]{Metzger2008}
{Metzger}, B.~D., {Piro}, A.~L., \& {Quataert}, E. 2008, \mnras, 390, 781

\bibitem[{{Metzger} {et~al.}(2018){Metzger}, {Thompson}, \&
  {Quataert}}]{Metzger2018}
{Metzger}, B.~D., {Thompson}, T.~A., \& {Quataert}, E. 2018, ArXiv e-prints,
  arXiv:1801.04286

\bibitem[{{Metzger} {et~al.}(2010){Metzger}, {Mart{\'{\i}}nez-Pinedo},
  {Darbha}, {Quataert}, {Arcones}, {Kasen}, {Thomas}, {Nugent}, {Panov}, \&
  {Zinner}}]{Metzger2010}
{Metzger}, B.~D., {Mart{\'{\i}}nez-Pinedo}, G., {Darbha}, S., {et~al.} 2010,
  \mnras, 406, 2650

\bibitem[{{Murguia-Berthier} {et~al.}(2014){Murguia-Berthier}, {Montes},
  {Ramirez-Ruiz}, {De Colle}, \& {Lee}}]{Murguia-Berthier2014}
{Murguia-Berthier}, A., {Montes}, G., {Ramirez-Ruiz}, E., {De Colle}, F., \&
  {Lee}, W.~H. 2014, \apjl, 788, L8

\bibitem[{{Murguia-Berthier} {et~al.}(2017){Murguia-Berthier}, {Ramirez-Ruiz},
  {Montes}, {De Colle}, {Rezzolla}, {Rosswog}, {Takami}, {Perego}, \&
  {Lee}}]{Murguia-Berthier2017}
{Murguia-Berthier}, A., {Ramirez-Ruiz}, E., {Montes}, G., {et~al.} 2017, \apjl,
  835, L34

\bibitem[{{Nakar} \& {Piran}(2017)}]{Nakar2017}
{Nakar}, E., \& {Piran}, T. 2017, \apj, 834, 28

\bibitem[{{Nakar} \& {Piro}(2014)}]{Nakar2014}
{Nakar}, E., \& {Piro}, A.~L. 2014, \apj, 788, 193

\bibitem[{{Nakar} \& {Sari}(2010)}]{Nakar2010}
{Nakar}, E., \& {Sari}, R. 2010, \apj, 725, 904

\bibitem[{{Pian} {et~al.}(2017){Pian}, {D'Avanzo}, {Benetti}, {Branchesi},
  {Brocato}, {Campana}, {Cappellaro}, {Covino}, {D'Elia}, {Fynbo}, {Getman},
  {Ghirlanda}, {Ghisellini}, {Grado}, {Greco}, {Hjorth}, {Kouveliotou},
  {Levan}, {Limatola}, {Malesani}, {Mazzali}, {Melandri}, {M{\o}ller},
  {Nicastro}, {Palazzi}, {Piranomonte}, {Rossi}, {Salafia}, {Selsing},
  {Stratta}, {Tanaka}, {Tanvir}, {Tomasella}, {Watson}, {Yang}, {Amati},
  {Antonelli}, {Ascenzi}, {Bernardini}, {Bo{\"e}r}, {Bufano}, {Bulgarelli},
  {Capaccioli}, {Casella}, {Castro-Tirado}, {Chassande-Mottin}, {Ciolfi},
  {Copperwheat}, {Dadina}, {De Cesare}, {di Paola}, {Fan}, {Gendre},
  {Giuffrida}, {Giunta}, {Hunt}, {Israel}, {Jin}, {Kasliwal}, {Klose}, {Lisi},
  {Longo}, {Maiorano}, {Mapelli}, {Masetti}, {Nava}, {Patricelli}, {Perley},
  {Pescalli}, {Piran}, {Possenti}, {Pulone}, {Razzano}, {Salvaterra},
  {Schipani}, {Spera}, {Stamerra}, {Stella}, {Tagliaferri}, {Testa}, {Troja},
  {Turatto}, {Vergani}, \& {Vergani}}]{Pian2017}
{Pian}, E., {D'Avanzo}, P., {Benetti}, S., {et~al.} 2017, \nat, 551, 67

\bibitem[{{Piro} \& {Kollmeier}(2017)}]{Piro2017}
{Piro}, A.~L., \& {Kollmeier}, J.~A. 2017, ArXiv e-prints, arXiv:1710.05822

\bibitem[{{Piro} \& {Nakar}(2013)}]{Piro2013}
{Piro}, A.~L., \& {Nakar}, E. 2013, \apj, 769, 67

\bibitem[{{Pozanenko} {et~al.}(2018){Pozanenko}, {Barkov}, {Minaev}, {Volnova},
  {Mazaeva}, {Moskvitin}, {Krugov}, {Samodurov}, {Loznikov}, \&
  {Lyutikov}}]{Pozanenko2018}
{Pozanenko}, A.~S., {Barkov}, M.~V., {Minaev}, P.~Y., {et~al.} 2018, \apjl,
  852, L30

\bibitem[{{Roberts} {et~al.}(2011){Roberts}, {Kasen}, {Lee}, \&
  {Ramirez-Ruiz}}]{Roberts2011}
{Roberts}, L.~F., {Kasen}, D., {Lee}, W.~H., \& {Ramirez-Ruiz}, E. 2011, \apjl,
  736, L21

\bibitem[{{Rosswog}(2005)}]{Rosswog2005}
{Rosswog}, S. 2005, \apj, 634, 1202

\bibitem[{{Schlafly} \& {Finkbeiner}(2011)}]{Schlafly2011}
{Schlafly}, E.~F., \& {Finkbeiner}, D.~P. 2011, \apj, 737, 103

\bibitem[{{Shappee} {et~al.}(2017){Shappee}, {Simon}, {Drout}, {Piro},
  {Morrell}, {Prieto}, {Kasen}, {Holoien}, {Kollmeier}, {Kelson}, {Coulter},
  {Foley}, {Kilpatrick}, {Siebert}, {Madore}, {Murguia-Berthier}, {Pan},
  {Prochaska}, {Ramirez-Ruiz}, {Rest}, {Adams}, {Alatalo}, {Ba{\~n}ados},
  {Baughman}, {Bernstein}, {Bitsakis}, {Boutsia}, {Bravo}, {Di Mille}, {Higgs},
  {Ji}, {Maravelias}, {Marshall}, {Placco}, {Prieto}, \& {Wan}}]{Shappee2017}
{Shappee}, B.~J., {Simon}, J.~D., {Drout}, M.~R., {et~al.} 2017, Science, 358,
  1574

\bibitem[{{Smartt} {et~al.}(2017){Smartt}, {Chen}, {Jerkstrand}, {Coughlin},
  {Kankare}, {Sim}, {Fraser}, {Inserra}, {Maguire}, {Chambers}, {Huber},
  {Kr{\"u}hler}, {Leloudas}, {Magee}, {Shingles}, {Smith}, {Young}, {Tonry},
  {Kotak}, {Gal-Yam}, {Lyman}, {Homan}, {Agliozzo}, {Anderson}, {Angus},
  {Ashall}, {Barbarino}, {Bauer}, {Berton}, {Botticella}, {Bulla}, {Bulger},
  {Cannizzaro}, {Cano}, {Cartier}, {Cikota}, {Clark}, {De Cia}, {Della Valle},
  {Denneau}, {Dennefeld}, {Dessart}, {Dimitriadis}, {Elias-Rosa}, {Firth},
  {Flewelling}, {Fl{\"o}rs}, {Franckowiak}, {Frohmaier}, {Galbany},
  {Gonz{\'a}lez-Gait{\'a}n}, {Greiner}, {Gromadzki}, {Guelbenzu},
  {Guti{\'e}rrez}, {Hamanowicz}, {Hanlon}, {Harmanen}, {Heintz}, {Heinze},
  {Hernandez}, {Hodgkin}, {Hook}, {Izzo}, {James}, {Jonker}, {Kerzendorf},
  {Klose}, {Kostrzewa-Rutkowska}, {Kowalski}, {Kromer}, {Kuncarayakti},
  {Lawrence}, {Lowe}, {Magnier}, {Manulis}, {Martin-Carrillo}, {Mattila},
  {McBrien}, {M{\"u}ller}, {Nordin}, {O'Neill}, {Onori}, {Palmerio},
  {Pastorello}, {Patat}, {Pignata}, {Podsiadlowski}, {Pumo}, {Prentice}, {Rau},
  {Razza}, {Rest}, {Reynolds}, {Roy}, {Ruiter}, {Rybicki}, {Salmon}, {Schady},
  {Schultz}, {Schweyer}, {Seitenzahl}, {Smith}, {Sollerman}, {Stalder},
  {Stubbs}, {Sullivan}, {Szegedi}, {Taddia}, {Taubenberger}, {Terreran}, {van
  Soelen}, {Vos}, {Wainscoat}, {Walton}, {Waters}, {Weiland}, {Willman},
  {Wiseman}, {Wright}, {Wyrzykowski}, \& {Yaron}}]{Smartt2017}
{Smartt}, S.~J., {Chen}, T.-W., {Jerkstrand}, A., {et~al.} 2017, \nat, 551, 75

\bibitem[{{STScI Development Team}(2013)}]{Pysynphot2013}
{STScI Development Team}. 2013, {pysynphot: Synthetic photometry software
  package}, Astrophysics Source Code Library, , , ascl:1303.023

\bibitem[{{Tanaka} \& {Hotokezaka}(2013)}]{Tanaka2013}
{Tanaka}, M., \& {Hotokezaka}, K. 2013, \apj, 775, 113

\bibitem[{{Tanvir} {et~al.}(2017){Tanvir}, {Levan},
  {Gonz{\'a}lez-Fern{\'a}ndez}, {Korobkin}, {Mandel}, {Rosswog}, {Hjorth},
  {D'Avanzo}, {Fruchter}, {Fryer}, {Kangas}, {Milvang-Jensen}, {Rosetti},
  {Steeghs}, {Wollaeger}, {Cano}, {Copperwheat}, {Covino}, {D'Elia}, {de Ugarte
  Postigo}, {Evans}, {Even}, {Fairhurst}, {Figuera Jaimes}, {Fontes}, {Fujii},
  {Fynbo}, {Gompertz}, {Greiner}, {Hodosan}, {Irwin}, {Jakobsson},
  {J{\o}rgensen}, {Kann}, {Lyman}, {Malesani}, {McMahon}, {Melandri},
  {O'Brien}, {Osborne}, {Palazzi}, {Perley}, {Pian}, {Piranomonte}, {Rabus},
  {Rol}, {Rowlinson}, {Schulze}, {Sutton}, {Th{\"o}ne}, {Ulaczyk}, {Watson},
  {Wiersema}, \& {Wijers}}]{Tanvir2017}
{Tanvir}, N.~R., {Levan}, A.~J., {Gonz{\'a}lez-Fern{\'a}ndez}, C., {et~al.}
  2017, \apjl, 848, L27

\bibitem[{{Troja} {et~al.}(2017){Troja}, {Piro}, {van Eerten}, {Wollaeger},
  {Im}, {Fox}, {Butler}, {Cenko}, {Sakamoto}, {Fryer}, {Ricci}, {Lien}, {Ryan},
  {Korobkin}, {Lee}, {Burgess}, {Lee}, {Watson}, {Choi}, {Covino}, {D'Avanzo},
  {Fontes}, {Gonz{\'a}lez}, {Khandrika}, {Kim}, {Kim}, {Lee}, {Lee}, {Kutyrev},
  {Lim}, {S{\'a}nchez-Ram{\'{\i}}rez}, {Veilleux}, {Wieringa}, \&
  {Yoon}}]{Troja2017}
{Troja}, E., {Piro}, L., {van Eerten}, H., {et~al.} 2017, \nat, 551, 71

\bibitem[{{Utsumi} {et~al.}(2017){Utsumi}, {Tanaka}, {Tominaga}, {Yoshida},
  {Barway}, {Nagayama}, {Zenko}, {Aoki}, {Fujiyoshi}, {Furusawa}, {Kawabata},
  {Koshida}, {Lee}, {Morokuma}, {Motohara}, {Nakata}, {Ohsawa}, {Ohta},
  {Okita}, {Tajitsu}, {Tanaka}, {Terai}, {Yasuda}, {Abe}, {Asakura}, {Bond},
  {Miyazaki}, {Sumi}, {Tristram}, {Honda}, {Itoh}, {Itoh}, {Kawabata},
  {Morihana}, {Nagashima}, {Nakaoka}, {Ohshima}, {Takahashi}, {Takayama},
  {Aoki}, {Baar}, {Doi}, {Finet}, {Kanda}, {Kawai}, {Kim}, {Kuroda}, {Liu},
  {Matsubayashi}, {Murata}, {Nagai}, {Saito}, {Saito}, {Sako}, {Sekiguchi},
  {Tamura}, {Tanaka}, {Uemura}, \& {Yamaguchi}}]{Utsumi2017}
{Utsumi}, Y., {Tanaka}, M., {Tominaga}, N., {et~al.} 2017, \pasj, 69, 101

\bibitem[{{Valenti} {et~al.}(2017){Valenti}, {Sand}, {Yang}, {Cappellaro},
  {Tartaglia}, {Corsi}, {Jha}, {Reichart}, {Haislip}, \&
  {Kouprianov}}]{Valenti2017}
{Valenti}, S., {Sand}, D.~J., {Yang}, S., {et~al.} 2017, \apjl, 848, L24

\bibitem[{{Van Der Walt} {et~al.}(2011){Van Der Walt}, {Colbert}, \&
  {Varoquaux}}]{VanDerWalt2011}
{Van Der Walt}, S., {Colbert}, S.~C., \& {Varoquaux}, G. 2011, ArXiv e-prints,
  arXiv:1102.1523

\bibitem[{{Villar} {et~al.}(2017){Villar}, {Guillochon}, {Berger}, {Metzger},
  {Cowperthwaite}, {Nicholl}, {Alexander}, {Blanchard}, {Chornock},
  {Eftekhari}, {Fong}, {Margutti}, \& {Williams}}]{Villar2017}
{Villar}, V.~A., {Guillochon}, J., {Berger}, E., {et~al.} 2017, \apjl, 851, L21

\bibitem[{{Waxman} {et~al.}(2017){Waxman}, {Ofek}, {Kushnir}, \&
  {Gal-Yam}}]{Waxman2017}
{Waxman}, E., {Ofek}, E., {Kushnir}, D., \& {Gal-Yam}, A. 2017, ArXiv e-prints,
  arXiv:1711.09638

\end{thebibliography}

\end{document}